\documentclass[12pt]{article}
\usepackage{graphicx}

\newcommand\pubnumber{}
\newcommand\pubdate{\today}

\def\lpnhe{LPNHE, CNRS/IN2P3, Sorbonne Universit\'e, \\ Universit\'e Paris Diderot, Paris 75252, France}

\def\Title#1{\begin{center} {\Large #1 } \end{center}}
\def\Author#1{\begin{center}{ \sc #1} \end{center}}
\def\Address#1{\begin{center}{ \it #1} \end{center}}

\newcommand\pubblock{\rightline{\begin{tabular}{l} \pubnumber\\
         \pubdate  \end{tabular}}}
\newenvironment{Abstract}{\begin{quotation}  }{\end{quotation}}
\newenvironment{Presented}{\begin{quotation} \begin{center} 
             PRESENTED AT\end{center}\bigskip 
      \begin{center}\begin{large}}{\end{large}\end{center} \end{quotation}}

\def\beq{\begin{equation}}
\def\eeq#1{\label{#1}\end{equation}}
\def\eeqn{\end{equation}}

\def\beqa{\begin{eqnarray}}
\def\eeqa#1{\label{#1}\end{eqnarray}}
\def\eeqan{\end{eqnarray}}

\let\bar=\overbar

\def\Dslash{\not{\hbox{\kern-4pt $D$}}}
\def\dslash{\not{\hbox{\kern-2pt $\del$}}}

\def\msb{{\bar{\ssstyle M \kern -1pt S}}}

\input symbols.tex

\begin{document}
\begin{titlepage}
\pubblock

\vfill
\Title{Latest results from T2K and T2K Phase II}
\vfill
\Author{ Claudio Giganti\footnote{cgiganti@lpnhe.in2p3.fr} \\
 (for the T2K collaboration)}
\Address{\lpnhe}
\vfill
\begin{Abstract}
T2K is a long-baseline neutrino oscillation experiment using a \num beam produced at the J-PARC facility. Neutrinos are detected at a Near Detector complex (ND280) and at the Far Detector (Super-Kamiokande). The most recent oscillation results presented by T2K are described in these proceedings. With these data T2K has produced the first constraints on \dcp, excluding CP conserving values at 2$\sigma$ and the world best measurement of the \thatm mixing angle.
\end{Abstract}
\vfill
\begin{Presented}
NuPhys2017, Prospects in Neutrino Physics \\

Barbican Centre, London, UK,  December 20--22, 2017\end{Presented}
\vfill
\end{titlepage}
\def\thefootnote{\fnsymbol{footnote}}
\setcounter{footnote}{0}

\section{Neutrino oscillations}
A large variety of experimental results using neutrinos from very different sources has contributed to establish the phenomenon of neutrino oscillations that are described within the PMNS~\cite{pmns} framework. The PMNS matrix is a 3$\times$3 unitary mixing matrix and is parametrized by three mixing angles, \thsol, \thatm, \thint, and a $CP$ violating phase, \dcp (two additional phases are present if neutrinos are Majorana particles but they do not affect oscillations). The additional parameters governing neutrino oscillations are the squared--mass differences $\Delta m_{ij}^2 = m_j^2 - m_i^2$, where $m_i$ is the mass of the $i$-th neutrino mass eigenstates. 

The original discovery of neutrino oscillations and the first measurements of  the corresponding mixing angles (\thsol and \thatm) and mass squared differences from \sk~\cite{sk}, SNO~\cite{sno}, and KamLAND~\cite{kamland} started a broad program in which neutrino oscillations have been observed by several experiments using very different neutrino sources and detection techniques. 

The last relevant milestones have been the discovery that also the last unknown mixing angle, \thint, is different from zero. After first indications from T2K in the   \num $\rightarrow$ \nue transition~\cite{t2k_2011}, \thint was measured to be different from zero in 2012 by Daya Bay~\cite{dayabay} and RENO~\cite{reno}. This discovery started the era of precision measurements of neutrino oscillations with the possibility of investigating sub-leading order effects to determine the mass ordering and to observe $CP$ violation in the leptonic sector. Such measurement is today the main goal of the T2K experiment.

\section{The T2K experiment}
T2K (Tokai to Kamioka)~\cite{t2knim} is a long-baseline neutrino oscillation experiment originally intended to measure \thint by observing electron neutrino appearance. A muon neutrino beam is produced at the J-PARC accelerator complex on the East Coast of Japan by striking a 30 GeV proton beam onto a 90-cm long carbon target. This produces hadrons that are focused and selected in charge by a system of magnetic horns and are directed towards a decay tunnel where they decay into neutrinos. By changing the direction of the current in the magnetic horns it is possible to select in charge different hadrons. If positively charged pions  are focused they decay into \mup and \num (\fhc) while if negatively charged pions are focused they decay into \mun and \numb (\rhc). The undecayed pions and other hadrons, as well as
the muons, are stopped by a beam dump, installed $100~m$ downstream the target.

\begin{figure}[htb]
\centering
\includegraphics[height=1.2in]{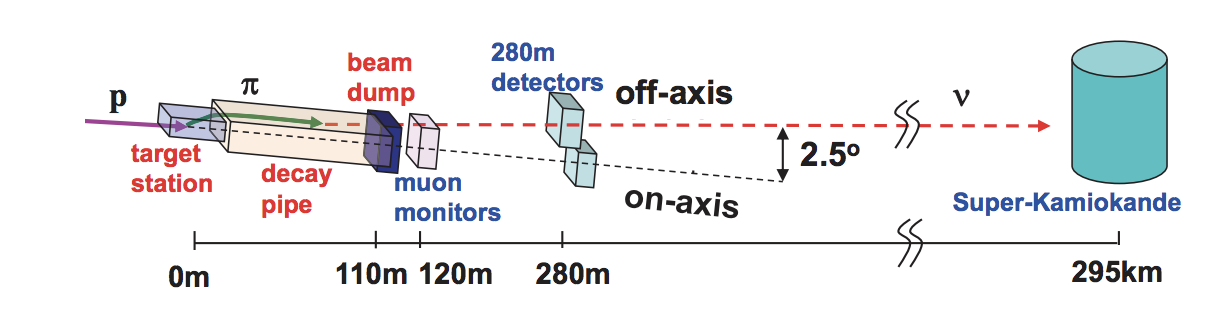}
\caption{A schematic view of the T2K neutrino beamline and detectors.}
\label{fig:t2kscheme}
\end{figure}

Neutrinos are then observed  in a Near Detector, ND280, at 280~m from the target, where the effect of the oscillations is negligible, and at the first oscillation peak at the far detector, \sk, 295 km away from J-PARC. The neutrino energy, peaked at 600~MeV, and the distance are  chosen to be at the expected maximum of the oscillations in order to maximize the sensitivity to \num  (\numb) disappearance and to \nue (\nueb) appearance. A schematic view of T2K is shown in Fig.~\ref{fig:t2kscheme}.

The near detector complex comprises an on-axis detector (INGRID) and an off-axis detector (ND280). INGRID is composed of 14 modules of iron and plastic scintillator spanning the neutrino beam in a transverse section of 10$\times$10~meters. Its goal is to measure, on a day--by--day basis the neutrino beam direction and profile.

\begin{figure}[htb]
\centering
\includegraphics[width=7cm]{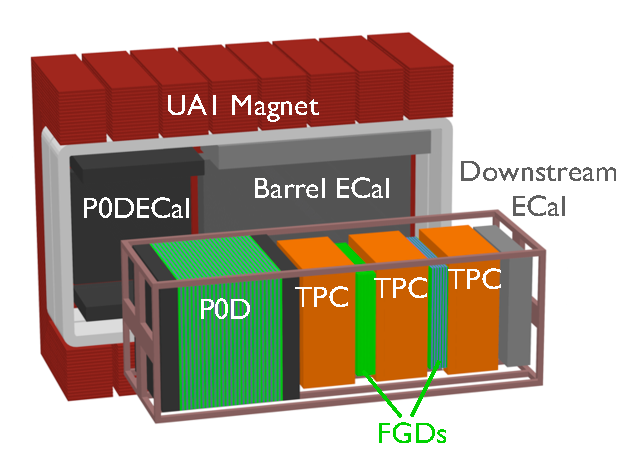}
\includegraphics[width=7cm]{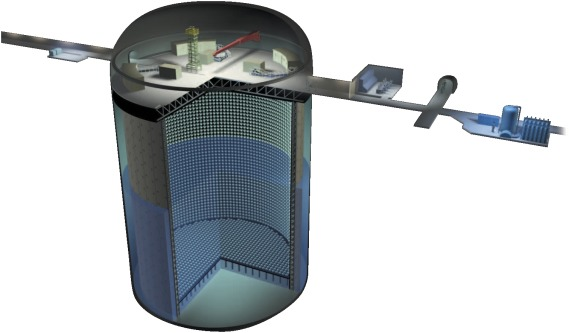}
\caption{A schematic view of ND280 (left) and \sk (right).}
\label{fig:nd280sk}
\end{figure}

The off-axis detector, ND280 (see Fig.~\ref{fig:nd280sk}), consists of several detectors installed in the
ex-UA1 magnet, operated at $0.2~T$: a $\pi^0$ detector (P0D) to
measure interactions with $\pi^0$ production, an electromagnetic
calorimeter (ECAL) to measure the electromagnetic activity and a Side
Muon Range Detector (SMRD) embedded in the magnet yokes. Finally a
Tracker system, composed of two Fine Grained Detectors (FGD) and three
Time Projection Chambers (TPC). 

Each FGD has a mass of $\sim1$~ton and acts as active target for the neutrino interactions.  The first FGD is a fully active detector, while in the second FGD scintillator layers are interleaved with inactive water layers, allowing to select neutrino interactions on carbon and on oxygen.
The three TPCs are used to do a 3D tracking of the charged particles produced in interactions in one of the FGDs, and to measure their charge and momentum from the curvature induced by the magnetic field. The particle identification is performed based on the measurement of the ionization.

The far detector of T2K is \sk, a 50~kton water Cherenkov detector located at a depth of 2700 meters water equivalent in the Kamioka mine (see Fig.~\ref{fig:nd280sk}).
\sk has a cylindrical shape with two concentric optically separated regions instrumented with Hamamatsu PMTs.  Neutrino interactions with water produce Cherenkov light which can be used to distinguish
between electron-like and muon-like events by analyzing the sharpness of a
Cherenkov ring. A muon makes a sharp edged ring whilst an electron makes a
fuzzy one due to electromagnetic showers. The electron/muon
misidentification probability, estimated using atmospheric neutrinos,
is about 1\% for the T2K neutrino energy. 

T2K has started the data taking in 2010 and, up to the Summer 2017, has collected 2.25\pot (protons--on--target), 1.49\pot in \fhc and 0.76\pot in \rhc. T2K is currently running in \rhc. 

\section{T2K oscillation analyses}

For the T2K oscillation analyses, the expected event rates and spectra at \sk 
are predicted based on a model of  neutrino fluxes and of neutrino 
cross-sections and measurements of neutrino interactions at ND280. More details on the oscillation analyses are given in~\cite{Abe:2017vif}. 

The flux modelling is based on the \shine hadroproduction 
measurements~\cite{Abgrall:2015hmv}, that allow reduction of the uncertainties on the fluxes below 10\%. 
The cross-section model is based on external measurements
from different experiments (mostly MiniBooNE and Minerva, see here for details~\cite{Wilkinson:2016wmz}). 
Uncertainties on event rates and spectra of the order of 15\% 
would be expected if only those data were available. 

\begin{figure} [h]
\includegraphics[width=7.5cm]{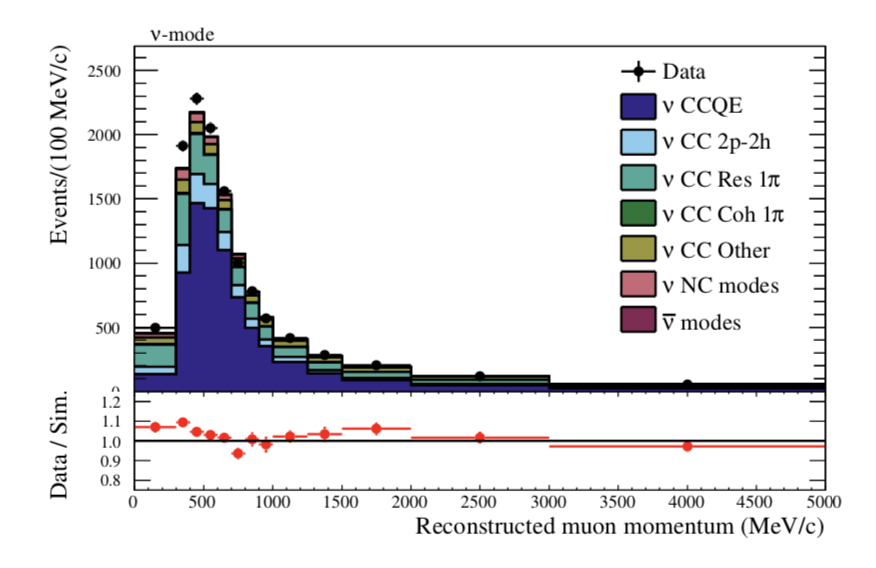}
\includegraphics[width=7.5cm]{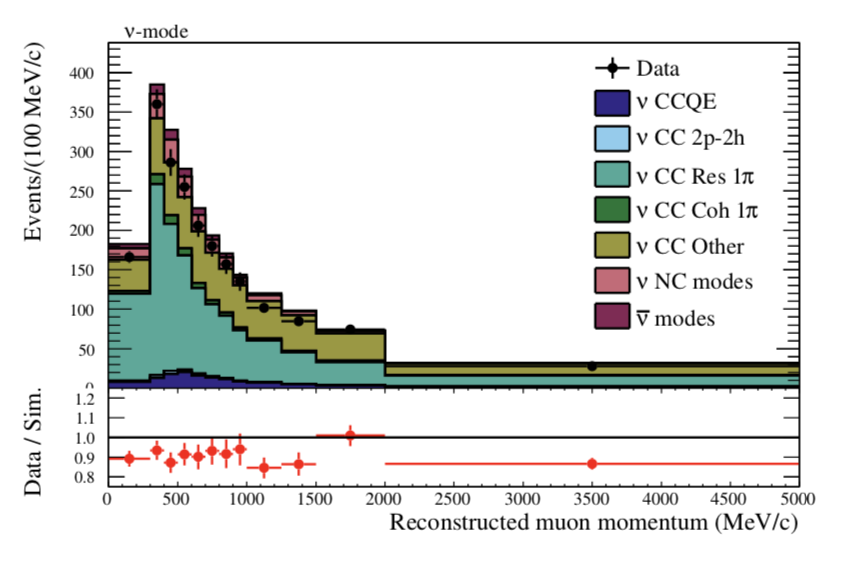}
\caption{\label{fig:nd280_banff}
Momentum distribution of outgoing muons for $\nu_\mu$ 
CC-0\pip (left) and CC-1\pip (right) samples at ND280.}
\end{figure}

Crucial inputs to the T2K oscillation analyses are then the measurements 
at the Near Detector. In the ND280 tracker, a total of 14 samples 
of \num and \numb charged current interactions are selected in the FGD1 
and in the FGD2 with muons precisely measured in the TPCs. 
The samples are separated according to the number of pions observed 
in the final state (0, 1, more than 1). Examples of these distributions 
as a function of outgoing muon momentum are shown in Fig.~\ref{fig:nd280_banff}. 

\begin{figure} [h]
\includegraphics[width=7.5cm]{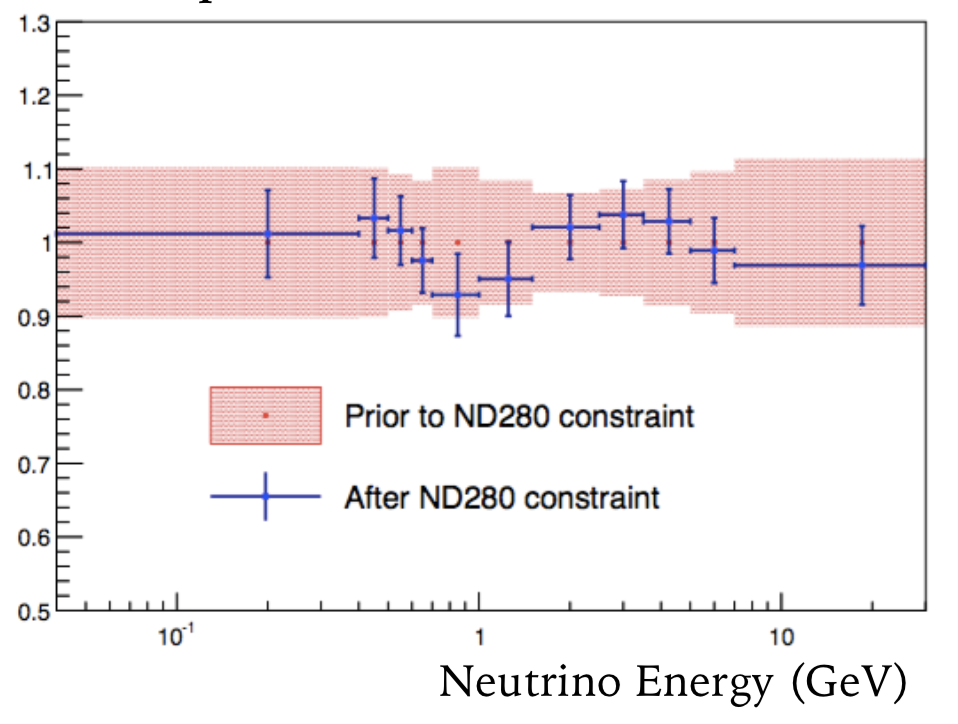}
\includegraphics[width=7.5cm]{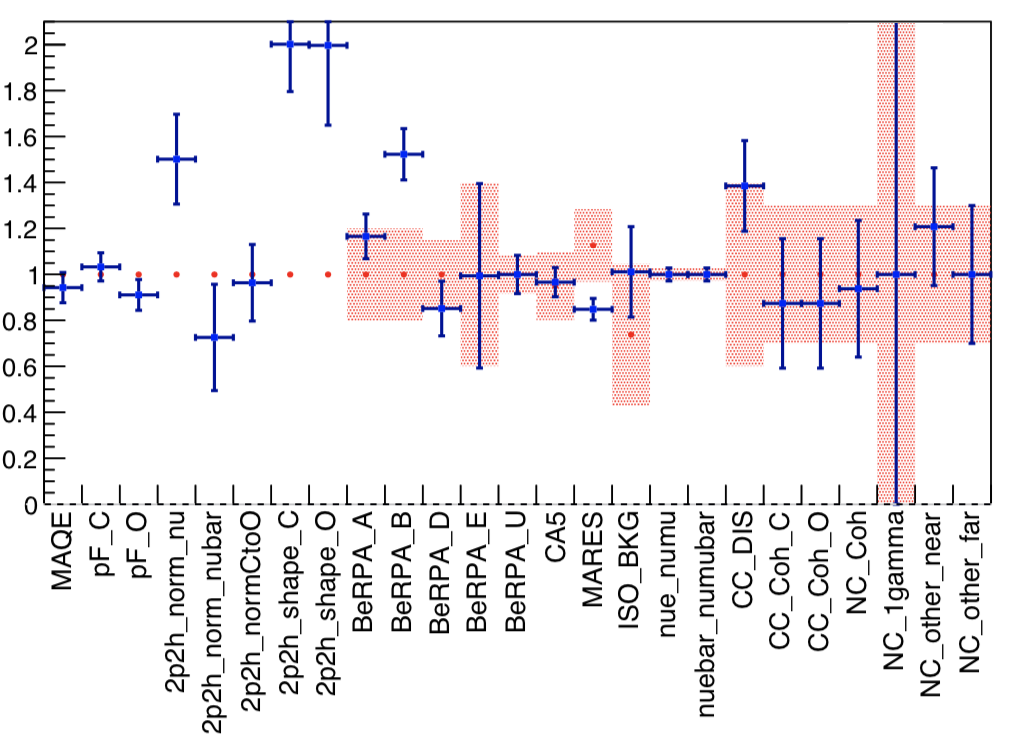}
\caption{\label{fig:banff_res}Flux (left) and cross-section (right) parameters before (red) and after (blue) the Near Detector fit.}
\end{figure}

The 14 event samples, selected in data and Monte Carlo, are binned in $p_{\mu}$ and $\cos\theta_{\mu}$ (where $\theta$ is the angle between the neutrino beam and the lepton candidate track) and fitted with a likelihood fit. 
The likelihood assumes that the observed number of events in each bin 
follows a Poisson distribution, with an expectation calculated according 
to the flux, cross-section and detector systematic parameters. The fitted neutrino cross-section and 
unoscillated SK flux parameters are passed to the oscillation analysis, 
using a covariance matrix to describe their uncertainties. 
The fit results for these parameters are shown in Fig.~\ref{fig:banff_res}. The flux parameters are well within the priors, while for the cross-section parameters the fit tends to increase the 2p2h component and the cross-section at low and intermediate values of $Q^2$ (BeRPA A and BeRPA B in the plot). A systematic uncertainty on the number of expected events at \sk in the range of $\sim$4--7\% is obtained as a result of this fit. 

A major improvement of the new T2K oscillation analysis with respect to previous analyses is that a new reconstruction algorithm is used for the \sk event selection. This algorithm combines time and charge likelihood for a given ring hypothesis. The better performance of this algorithm allows for a new definition of the Fiducial Volume in which not only the distance of the vertex from the wall but also the direction of the lepton candidate with respect to the wall is used. 

This new \sk selection and the new definition of the fiducial volume, allow to increase by 30\% the efficiency in selecting e--like samples while keeping the same purity of $\sim80\%$. For the $\mu$--like sample the new selection allows to increase the purity in selecting charged--current interactions without pions in the final state from 70\% to 80\%.

\begin{figure}[!h]
\begin{center}
\includegraphics[width=4.5cm]{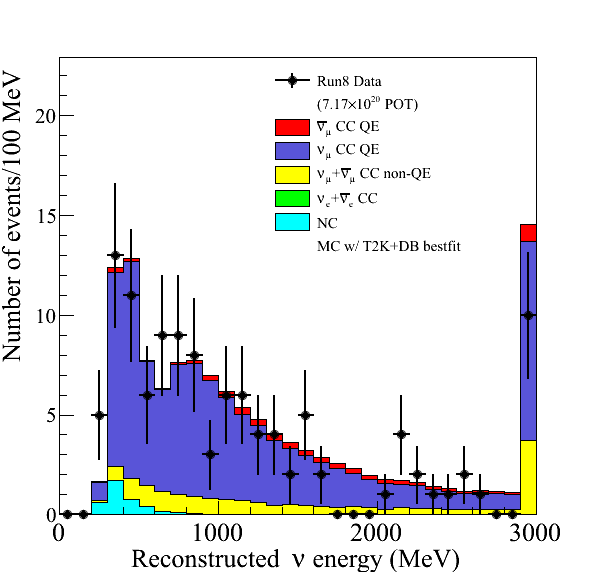}
\includegraphics[width=4.5cm]{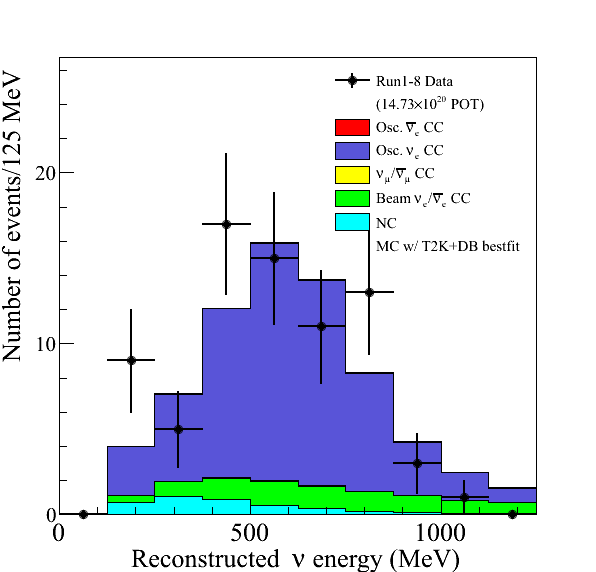}
\includegraphics[width=4.5cm]{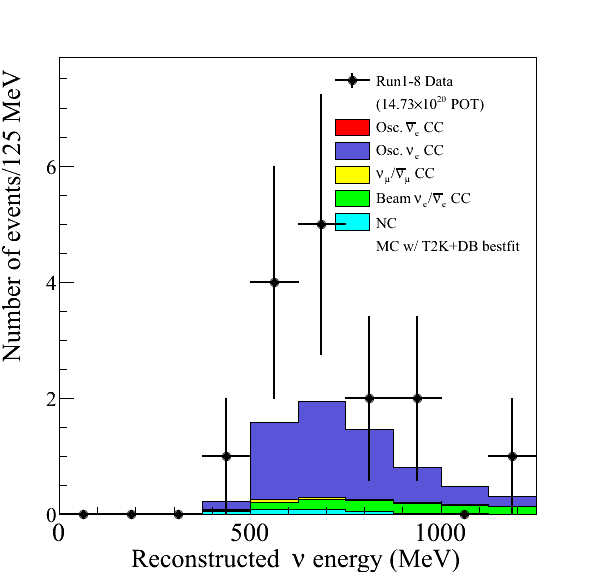}
\includegraphics[width=4.cm]{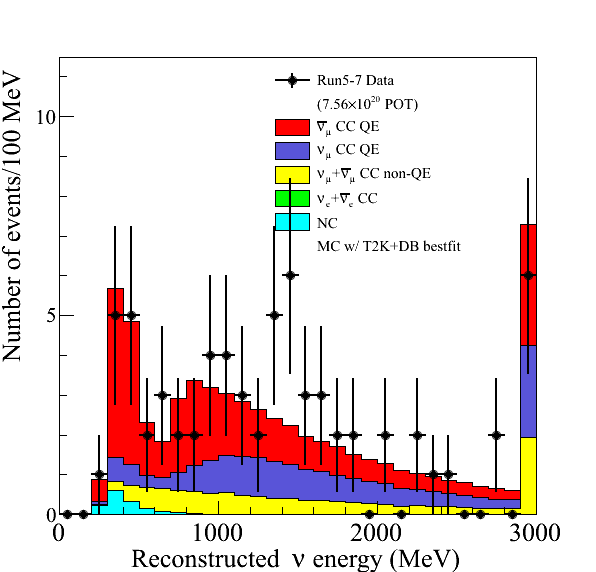}
\includegraphics[width=4.cm]{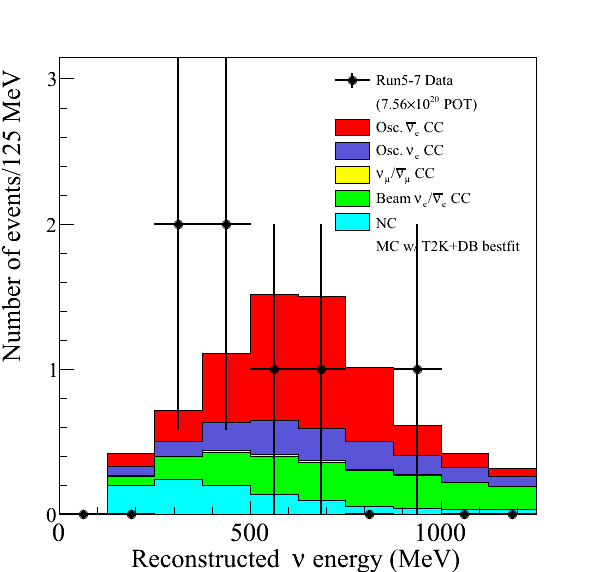}
\caption{\label{fig:sk_spectra}Observed spectra at SK in the five samples used for the oscillation analyses. Top: $\mu$--like, e-like, e--like+1\pipi in \fhc. Bottom: $\mu$--like and e--like in \rhc.}
\end{center}
\end{figure}

Five samples are selected at \sk and are used in the oscillation analyses: single--ring $\mu$--like events selected in \fhc and in \rhc, single--ring e--like events selected in \fhc and in \rhc, and a fifth sample, selected only in \fhc, where the e--like ring is accompanied by the presence of a delayed electron, due to the decay of a pion produced in the neutrino interaction. The number of events selected at \sk in the 5 samples are presented 
in Tab.~\ref{tab:sk_selected} and compared with the expected numbers 
of events for different values of \dcp. The spectra are shown 
in Fig.~\ref{fig:sk_spectra}. 

\begin{table} [!h]
  \centering
  \begin{tabular}{| c | c | c | c | c | c | }
  \hline
 & {\bf Data} & MC & MC & MC & MC \\
 & & (\dcp = -\pipi/2) & (\dcp = 0) &  (\dcp = \pipi/2) &  (\dcp = \pipi)\\
    \hline
e--like \fhc & {\bf 74} & 73.5 & 61.5 & 49.9 & 61.5 \\
 \hline
e--like + 1\pipi \fhc & {\bf 15} & 6.9 & 6.0 & 4.9 & 5.8 \\
 \hline
e--like \rhc & {\bf 7} & 7.9 & 9.0 & 10.0 & 8.9 \\
 \hline
$\mu$--like \fhc & {\bf 240} & 267.8 & 267.4 & 267.7 & 268.2 \\
\hline
$\mu$--like \rhc & {\bf 68} & 63.1 & 62.9 & 63.1 & 63.1 \\
\hline
\end{tabular}
\caption{\label{tab:sk_selected} Observed and expected numbers of events at SK for different values of \dcp.}
\end{table}

As it is clear from Tab.~\ref{tab:sk_selected}, 
\dcp only affects the e-like samples and values of \dcp close 
to -\pipi/2 tends to increase the \nue appearance probability, 
while decreasing the \nueb probability. This is exactly what is observed 
in the data in \fhc (\rhc), where 74 (7) single-ring e-like events 
are observed while 62 (9) are expected if \dcp = 0 or \pipi.

\begin{figure}[!h]
\begin{center}
\includegraphics[width=7.5cm]{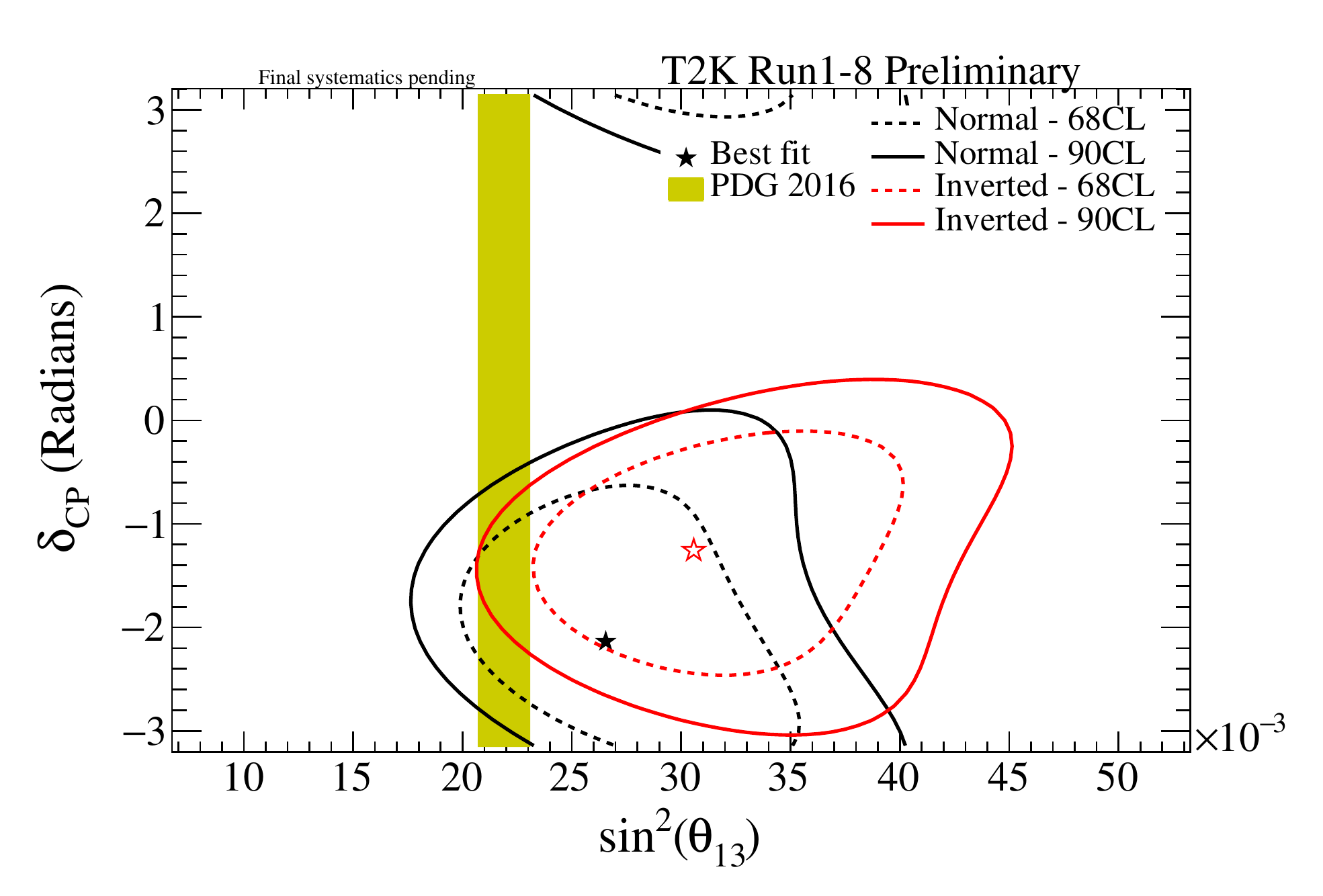}
\includegraphics[width=7.5cm]{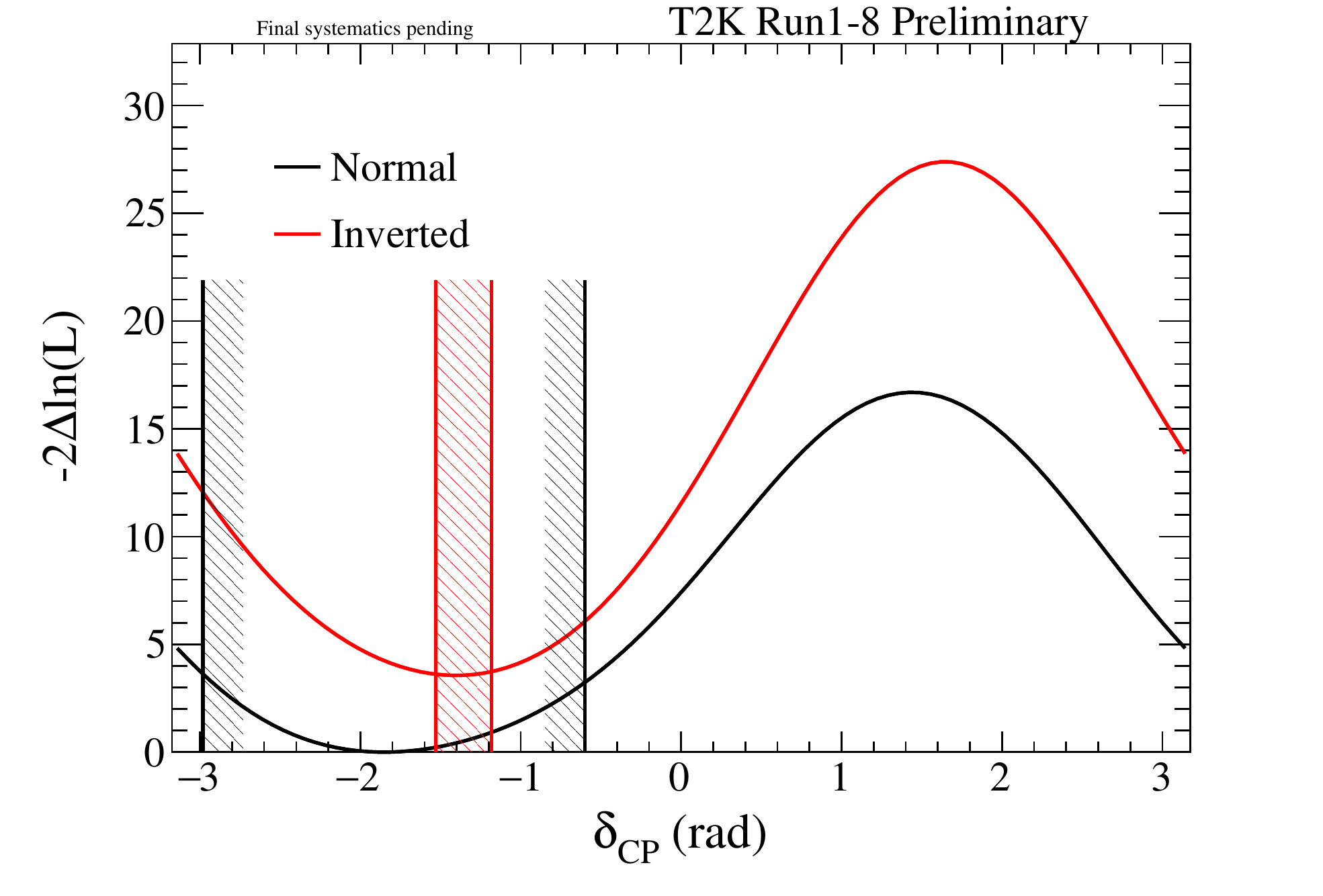}
\caption{\label{fig:t2koaresdcp}
Measurement of the oscillation parameters \thint and \dcp without  reactor constraint and measurement of \dcp with reactor constraints. The bands on the right plot represent the 95\% CL allowed regions for the two hierarchies.}
\end{center}
\end{figure}

\begin{figure} [!h]
\begin{center}
\includegraphics[width=7.5cm]{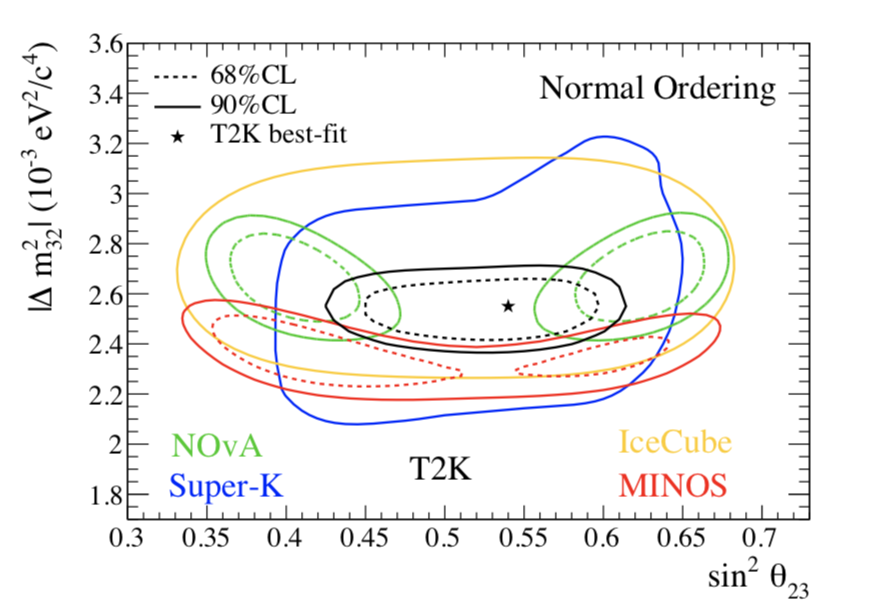}
\caption{\label{fig:t2koaresatm}Measurement of the oscillation parameters \thatm and \dmsq from T2K, compared to other experiments.}
\end{center}
\end{figure}

The five samples
are then fitted together in order to extract 
the oscillation parameters \thatm, \dmsq, \thint, and \dcp. 
The value of \thint can either be a free parameter in the fit 
or it can be constrained to the precise measurement 
of the reactor experiments. The two cases are shown 
in Fig.~\ref{fig:t2koaresdcp}: both fits prefer values 
of \dcp close to -\pipi/2 and, when the reactor constraint 
is included, the CP conserving values 0 and \pipi are excluded 
at more than 95\% CL. \thatm and \dmsq are also precisely determined 
by T2K: in particular the value of \thatm is  compatible 
with maximal mixing as shown in Fig.~\ref{fig:t2koaresatm}. 
It should be noticed that in Fig.~\ref{fig:t2koaresatm} some tensions 
are observed between T2K and the first published \nova results 
for \thatm~\cite{Adamson:2017qqn}. 
In a recent update of the oscillation analysis of \nova, 
\thatm is found to be compatible with maximal mixing 
and currently there is no tension 
between T2K and \nova results.

\section{T2K phase II}
\begin{figure} [h!]
\begin{center}
\includegraphics[width=7.5cm]{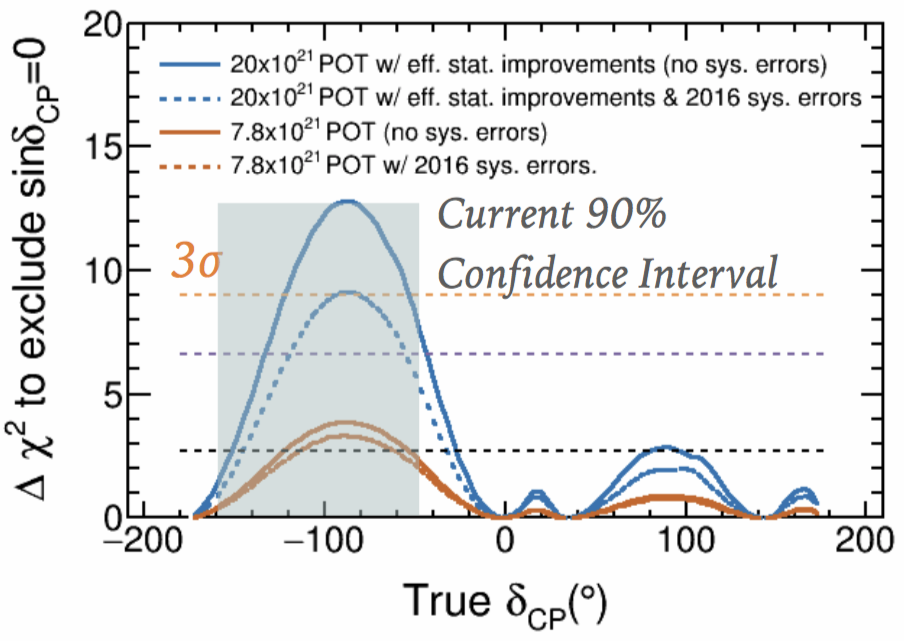}
\includegraphics[width=7.5cm]{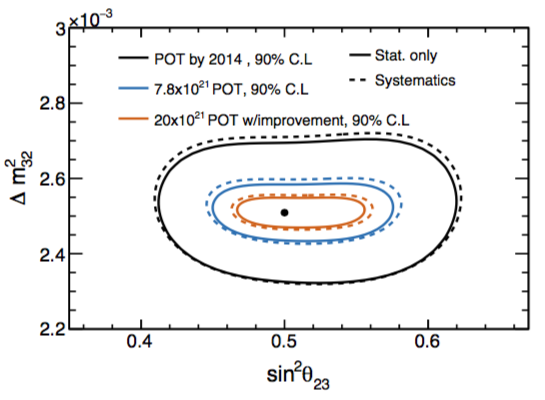}
\caption{\label{fig:t2ksensi} Left: Expected T2K-II sensitivity to \dcp assuming the mass ordering is not known as a function of \dcp. Right: Expected T2K-II sensitivity to \thatm and \dmsq.}
\end{center}
\end{figure}

T2K was originally approved to collect 7.8\pot and such statistics was mainly driven by the sensitivity of the experiment to \thint, corresponding to the value at which systematic uncertainties  would have been dominant in case of small \thint. Today we know that \thint is large and an increased statistics would improve the sensitivity of the experiment to measure \dcp as shown in Fig.~\ref{fig:t2ksensi}. For this reason T2K has proposed an extension of the running time that will allow to collect  a statistics of  20\pot. By collecting this statistics T2K will be able to observe CP violation with more than 3$\sigma$ significance if CP violation is large and to measure  $\theta_{23}$ and $\Delta m_{32}^{2}$, with a precision of 1.7$^\circ$ or better and 1\%, respectively. Such statistics will be obtained also thanks to an upgrade of the J-PARC Main Ring power supplies that will allow to reach $\sim$1~MW of beam  power (while currently T2K is stably running at $\sim470$~kW of beam power).

In order to fully profit from the foreseen additional statistics a better understanding on systematic uncertainties will be necessary. For this reason the T2K collaboration has launched an upgrade project for the Near Detector, aimed at overcoming the known limitations of the current design of ND280, that concerns the angular acceptance of the near and far detectors.

Thanks to the cylindrical shape of the tank and to its large size, in fact, \sk has an efficiency in selecting neutrino interactions that is independent on the lepton direction. The geometrical configuration of the ND280 tracker, instead, allows to select with excellent efficiency  tracks emitted parallel to the beam  but this efficiency rapidly degrades with the angle with respect to the beam, being close to zero for \costheta $\leq$ 0.4 (where $\theta$ is the angle between the emitted lepton and the beam).

The baseline proposal for the upgrade, which achieves  a much better uniformity of acceptance as function of polar angle, 
 includes a fully active scintillator detector 
 acting as neutrino target, disposed along the plane including both 
 the beam direction and the magnetic field. The favoured option 
 for this detector is the Super-FGD concept~\cite{superfgd}, consisting of small 
 scintillator cubes each read-out by three wave--length shifting fibers. 
 Two new TPCs, with concept similar to the one of the three existing TPCs, will be installed above and below the Super-FGD covering the large polar angle regions. Time-of-flight detectors will also be used to reject out of fiducial volume events. The goal is to install these new detectors in ND280 by 2021.

\end{document}